%% file: _Matthew_Ames__Innovations_in_Risk_Management__Revised_.tex
\documentclass[graybox]{svmult}

\usepackage{mathptmx}       
\usepackage{helvet}         
\usepackage{courier}        
\usepackage{type1cm}        
\usepackage{makeidx}         
\usepackage{graphicx}        
\usepackage{multicol}        
\usepackage[bottom]{footmisc}

\usepackage{placeins} 
\usepackage{booktabs} 

\usepackage{epstopdf} 

\usepackage{lscape} 

\usepackage{amsmath} 
\usepackage{amsfonts}
\usepackage{amssymb}
\usepackage{bm}

\def\E{{\mathbb E}}

\newtheorem{theo}{Theorem}[section]

\newtheorem{rems}[theo]{Remarks}

\newtheorem{model ass}[theo]{Model Assumptions}

\usepackage{multirow}
\usepackage{rotating}

\makeindex             

\begin{document}

\title*{Upside and Downside Risk Exposures of Currency Carry Trades via Tail Dependence}
\author{Matthew Ames, Gareth W. Peters, Guillaume Bagnarosa and Ioannis Kosmidis}
\institute{Matthew Ames \at Department of Statistical Science, University College London, UK,\newline \email{m.ames.12@ucl.ac.uk}
\and Gareth W. Peters \at Lecturer, Department of Statistical Science, University College London, UK;\newline Adjunct Scientist, Commonwealth Scientific and Industrial Research Organisation, Australia;
\newline Research Associate, Oxford-Man Institute, Oxford University, UK 
\and Guillaume Bagnarosa \at Assistant Professor, ESC Rennes School of Business; \newline
 Honorary Research Associate, Department of Computer Science, University College London, UK 
\and Ioannis Kosmidis \at Lecturer, Department of Statistical Science, University College London, UK 
}
%
%
\maketitle


\vspace{-3cm}

\abstract{Currency carry trade is the investment strategy that involves selling low interest rate currencies in order to purchase higher interest rate currencies, thus profiting from the interest rate differentials. This is a well known financial puzzle to explain, since assuming foreign exchange risk is uninhibited and the markets have rational risk-neutral investors, then one would not expect profits from such strategies. That is, according to uncovered interest rate parity (UIP), changes in the related exchange rates should offset the potential to profit from such interest rate differentials. However, it has been shown empirically, that investors can earn profits on average by borrowing in a country with a lower interest rate, exchanging for foreign currency, and investing in a foreign country with a higher interest rate, whilst allowing for any losses from exchanging back to their domestic currency at maturity. \newline
\indent This paper explores the financial risk that trading strategies seeking to exploit a violation of the UIP condition are exposed to with respect to multivariate tail dependence present in both the funding and investment currency baskets. It will outline in what contexts these portfolio risk exposures will benefit accumulated portfolio returns and under what conditions such tail exposures will reduce portfolio returns.
%
%
 }  



\input{Introduction_GWP_26_Feb_2014} 

\input{Joint_Tail_Risk_Exposure_GWP_26_02_2014_v1}


\input{FittingtheModels_GWP_26_Feb_2014_v1_0_0}

\input{DataDescriptionBasketFormation_GWP_27_Feb_2014_v1_0_0}

\input{Results_GWP_27_Feb_2014_v1}

\input{Discussion_MA_14_Nov_2013_v3_0_0}
\bibliographystyle{abbrv}
\bibliography{CFE_COPULA_Bibliography}

\end{document}

%% file: Introduction_GWP_26_Feb_2014.tex
\section{Currency Carry Trade and Uncovered Interest Rate Parity}

One of the most robust puzzles in finance still to be satisfactorily explained is the uncovered interest rate parity puzzle and the associated excess average returns of currency carry trade strategies. Such trading strategies are popular approaches which involve constructing portfolios by selling low interest rate currencies in order to buy higher interest rate currencies, thus profiting from the interest rate differentials. The presence of such profit opportunities, pointed out by \cite{Hansen1980,Fama1984,backus2001affine} and more recently by \cite{Lustig2007,Brunnermeier2008,burnside2011peso,christiansen2011time,Lustig2011,Menkhoff2012}, violates the fundamental relationship of uncovered interest rate parity (UIP). The UIP refers to the parity condition in which exposure to foreign exchange risk, with unanticipated changes in exchange rates, is uninhibited and therefore if one assumes rational risk-neutral investors, then changes in the exchange rates should offset the potential to profit from the interest rate differentials between high interest rate (investment) currencies and low interest rate (funding) currencies.
We can more formally write this relation by assuming that the forward price, $F_{t}^{T}$, is a martingale under the risk neutral probability $\mathbb{Q}$ (\cite{musiela2011martingale}): 
\begin{align}
E_{\mathbb{Q}}\Bigg[\frac{S_{T}}{S_{t}}\Bigg|\mathcal{F}_{t}\Bigg]=\frac{F_{t}^{T}}{S_{t}}=e^{(r_{t}-r_{t}^{\star})(T-t)}. \label{UIP}
\end{align}
The UIP Equation~(\ref{UIP}) thus states that under the risk neutral probability the expected variation of the exchange rate $S_{t}$ should equal the differential between the interest rate of the two associated countries, denoted by respectively $r_{t}$ and $r_{t}^{\star}$. The currency carry trade strategy investigated in this paper aims at exploiting violations of the UIP relation by investing a certain amount in a basket of high interest rate currencies (the long basket) while funding it through a basket of low interest rate currencies (the short basket).  When the UIP holds, then given foreign exchange market equilibrium, no profit should arise on average from this strategy; however such opportunities are routinely observed and exploited by large volume trading strategies.

In this paper we build on the existing literature by studying a stochastic feature of the joint tail behaviours of the currencies within each of the long and the short baskets, which form the carry trade. We aim to explore to what extent one can attribute the excess average returns with regard to compensation for exposure to tail risk, for example either dramatic depreciations in the value of the high interest rate currencies or dramatic appreciations in the value of the low interest rate currencies in times of high market volatility.

We postulate that such analyses should also benefit from consideration not only of the marginal behaviours of the processes under study, in this case the exchange rates of currencies in a portfolio, but also a rigorous analysis of the joint dependence features of such relationships. We investigate such joint relationships in light of the UIP condition. To achieve this, we study the probability of joint extreme movements in the funding and investment currency baskets and interpret these extremal tail probabilities as relative risk exposures of adverse and beneficial joint currency movements which would affect the portfolio returns. This allows us to obtain a relative contribution to the exposure of the portfolio profit decomposed in terms of the downside and upside risks that are contributed from such tail dependence features in each currency basket. We argue that the analysis of the carry trade is better informed by jointly modelling the multivariate behaviour of the marginal processes of currency baskets accounting for potential multivariate extremes, whilst still incorporating heavy-tailed relationships studied in marginal processes.


%
We fit mixture copula models to vectors of daily exchange rate log returns between 1989 - 2014 for both the investment and funding currency baskets making up the carry trade portfolio. The method and the dataset considered for the construction of the respective funding and investing currencies baskets are thoroughly described in \cite{Ames2013}.
The currency compositions of the funding and investment baskets are varying daily over time as a function of the interest rate differential processes for each currency relative to the USD.

  
Our analysis concludes that the appealing high return profile of a carry portfolio is not only compensating the tail thickness of each individual component probability distribution but also the fact that extreme returns tend to occur simultaneously and lead to a portfolio particularly sensitive to the risk of what is known as drawdown. Furthermore, we also demonstrate that high interest rate currency baskets and low interest rate currency baskets can display periods during which the tail dependence gets inverted, demonstrating when periods of construction of the aforementioned carry positions are being undertaken by investors.

%% file: Joint_Tail_Risk_Exposure_GWP_26_02_2014_v1.tex
\section{Interpreting Tail Dependence as Financial Risk Exposure in Carry Trade Portfolios}
\label{joint_tail_risk_exposure}

In order to fully understand the tail risks of joint exchange rate movements present when one invests in a carry trade strategy we can look at both the downside extremal tail exposure and the upside extremal tail exposure within the funding and investment baskets that comprise the carry portfolio. The downside tail exposure can be seen as the crash risk of the basket, i.e. the risk that one will suffer large joint losses from each of the currencies in the basket. These losses would be the result of joint appreciations of the currencies one is short in the low interest rate basket and/or joint depreciations of the currencies one is long in the high interest rate basket.

\begin{definition}[Downside Tail Risk Exposure in Carry Trade Portfolios]
\\Consider the funding currency (short) basket with $n$-exchange rates relative to base currency, on day $t$, with currency log-returns 
{\scriptsize{
$\smash{(X^{(1)}_t,X^{(2)}_t,\ldots,X^{(n)}_t)}$.}}
Then the downside tail exposure risk for the carry trade will be defined as the conditional probability of adverse currency movements in the short basket, corresponding to its upper tail dependence, given by
{\scriptsize{
\begin{equation}\label{EqnDown1}
\lambda_u^{(i)} (u) := \mathbb{P}\text{r}\left(X^{(i)}_t > F_i^{-1}(u)|X^{(1)}_t > F_1^{-1}(u),\ldots,X^{(i-1)}_t>F_{i-1}^{-1}(u), X^{(i+1)}_t>F_{i+1}^{-1}(u),\ldots,X^{(n)}_t>F_n^{-1}(u)\right)
\end{equation}
}}
for a currency of interest $i \in \left\{1,2,\ldots,n\right\}$. Conversely the downside tail exposure for the investment (long) basket with $n$ currencies will be defined as the conditional probability of adverse currency movement in the long basket, given by
{\scriptsize{
\begin{equation}\label{EqnDown2}
\lambda_l^{(i)}(u) := \mathbb{P}\text{r}\left(X^{(i)}_t < F_i^{-1}(u)|X^{(1)}_t < F_1^{-1}(u),\ldots,X^{(i-1)}_t < F_{i-1}^{-1}(u), X^{(i+1)}_t < F_{i+1}^{-1}(u),\ldots,X^{(n)}_t < F_n^{-1}(u)\right).
\end{equation}
}}
In general then a basket's upside or downside risk exposure would be quantified by the probability of a loss (or gain) arising from an appreciation or depreciation jointly of magnitude $u$ and the dollar cost associated to a given loss/gain of this magnitude. The standard approach in economics would be to associate say a linear cost function in $u$ to such a probability of loss to get say the downside risk exposure in dollars according to $E(u) = C_u({F_{X_t^{(i)}}(u)}) \times \lambda_u(u)$, which will be a function of the level $u$. As $\lambda_u$ becomes independent of the marginals, i.e. as $u \rightarrow 0$ or $u \rightarrow 1$, $C$ also becomes independent of the marginals.
\end{definition}

Conversely, we will also define the upside tail exposure that will contribute to profitable returns in the carry trade strategy when extreme movements that are in favour of the carry position held. These would correspond to precisely the probabilities discussed above applied in the opposite direction. That is the upside risk exposure in the funding (short) basket is given by Equation~(\ref{EqnDown1}) and the upside risk exposure in the investment (long) basket is given by Equation~(\ref{EqnDown2}). That is the upside tail exposure of the carry trade strategy is defined to be the risk that one will earn large joint profits from each of the currencies in the basket. These profits would be the result of joint depreciations of the currencies one is short in the low interest rate basket and/or joint appreciations of the currencies one is long in the high interest rate basket. 

\begin{rems} In a basket with $n$ currencies, $n \geq 2$, if one considers capturing the upside and downside financial risk exposures from a model based calculation of these extreme probabilities then
%
if the parametric model is exchangeable, such as an Archimedean copula, then swapping currency $i$ in Equation (\ref{EqnDown1}) and Equation (\ref{EqnDown2}) with another currency from the basket, say $j$ will not alter the downside or upside risk exposures. If they are not exchangeable then one can consider upside and downside risks for each individual currency in the carry trade portfolio.

%
%
\end{rems}

We thus consider these tail upside and downside exposures of the carry trade strategy as features that can show that even though average profits may be made from the violation of UIP, it comes at significant tail exposure. 

We can formalise the notion of the dependence behaviour in the extremes of the multivariate distribution through the concept of tail dependence, limiting behaviour of Equations (\ref{EqnDown1}) and (\ref{EqnDown2}), as $u \uparrow 1$ and $u \downarrow 0$ asymptotically.
The interpretation of such quantities is then directly relevant to assessing the chance of large adverse movements in multiple currencies which could potentially increase the risk associated with currency carry trade strategies significantly, compared to risk measures which only consider the marginal behaviour in each individual currency. 
Under certain statistical dependence models these extreme upside and downside tail exposures can be obtained analytically. we develop a flexible copula mixture example that has such properties below.



\section{Generalised Archimedean Copula Models for Currency Exchange Rate Baskets}

In order to study the joint tail dependence in the investment or funding basket we consider an overall tail dependence analysis which is parametric model based, obtained by using flexible mixtures of Archimedean copula components. Such a model approach is reasonable since typically the number of currencies in each of the long basket (investment currencies) and the short basket (funding currencies) is 4 or 5.

In addition these models have the advantage that they produce asymmetric dependence relationships in the upper tails and the lower tails in the multivariate model.
We consider three models; two Archimedean mixture models and one outer power transformed Clayton copula. The mixture models considered are the Clayton-Gumbel mixture and the Clayton-Frank-Gumbel mixture, where the Frank component allows for periods of no tail dependence within the basket as well as negative dependence. We fit these copula models to each of the long and short baskets separately.

\begin{definition}[Mixture Copula] 
A mixture copula is a linear weighted combination of copulae of the form:
\begin{equation}
C_M(\mathbf{u}; \mathbf{\theta}) = \sum_{i=1}^N \lambda_i C_i ({\bf u};{\bf \theta_i}),		
\end{equation}
where  $0 \leq \lambda_i \leq 1 \;\; \forall i \in \{1, ..., N\}$ and $\sum_{i=1}^N \lambda_i = 1$.
\end{definition}




\begin{definition}[Archimedean Copula]
A d-dimensional copula C is called Archim- edean if it can be represented by the form:
\begin{equation}
C({\bf u}) = \psi \{\psi^{-1}(u_1) + \cdots + \psi^{-1}(u_d)\} = \psi \{t({\bf u}) \} \;\;\;\; \forall {\bf u} = \{u_1, \ldots, u_d\} \in [0, 1]^d ,
\label{eq:archimedean_copula}
\end{equation}
where $\psi$ is an Archimedean generator satisfying the conditions given in \cite{McNeil2009}. $\psi ^{-1}:[0, 1] \rightarrow [0, \infty)$ is the inverse generator with $\psi^{-1}(0) = \mbox{inf}\{t: \psi(t) = 0\}$.
\end{definition}



In the following section we consider two stages to estimate the multivariate basket returns, firstly the estimation of suitable heavy tailed marginal models for the currency exchange rates (relative to USD), followed by the estimation of the dependence structure of the multivariate model composed of multiple exchange rates in currency baskets for long and short positions. 

Once the parametric Archimedean mixture copula model has been fitted to a basket of currencies, it is possible to obtain the upper and lower tail dependence coefficients, via closed form expressions for the class of mixture copula models and outer-power transform models we consider. The tail dependence expressions for many common bivariate copulae can be found in \cite{Nelsen2006}. This concept was recently extended to the multivariate setting by \cite{de2012multivariate}.

\begin{definition}[Generalized Archimedean Tail Dependence Coefficient] 
Let $X = (X_1,..., X_n)^T$ be an n-dimensional random vector with distribution \newline $C(F_1(X_1), \ldots, F_n(X_n))$, where $C$ is an Archimedean copula and $F_1, ..., F_n$ are the marginal distributions. The coefficients of upper and lower tail dependence are defined respectively as:
\small{
\begin{equation}
\begin{aligned}
\lambda_u^{1,...,h|h+1,...,n} &= \lim_{u \rightarrow 1-} P\left( X_1 > F_1^{-1}(u),...,X_h > F_h^{-1}(u) | X_{h+1} > F_{h+1}^{-1}(u), ..., X_n > F_n^{-1}(u) \right) \\
&= \lim_{t \rightarrow 0^+} \frac{\sum_{i=1}^d \left( \binom{d}{d-i} i (-1)^{i} \left[ \psi^{'}  (it) \right] \right) }{\sum_{i=1}^{n-h} \left( \binom{n-h}{n-h-i} i (-1)^i \left[ \psi^{'} (it) \right] \right)} \;\;\;\; ,
\end{aligned}
\label{eq:archmuppertd}
\end{equation}}
\begin{equation}
\begin{aligned}
\lambda_l^{1,...,h|h+1,...,n} &= \lim_{u \rightarrow 0+} P \left( X_1 < F_1^{-1}(u),...,X_h < F_h^{-1}(u) | X_{h+1} < F_{h+1}^{-1}(u), ..., X_n < F_n^{-1}(u) \right) \\
&= \lim_{t \rightarrow \infty} \frac{n}{n-h} \frac{\psi^{'} (nt)}{\psi^{'} ((n-h)t)}
\end{aligned}
\label{eq:archmlowertd}
\end{equation}
\normalsize
for the model dependence function `generator' $\psi(\cdot)$ and its inverse function.
\end{definition}



In \cite{de2012multivariate} the analogous form of the generalized multivariate upper and lower tail dependence coefficients for outer-power transformed Clayton copula models is provided.
The derivation of Equations (\ref{eq:archmuppertd}) and (\ref{eq:archmlowertd}) for the outer power case follows from \cite{feller1971}, i.e. the composition of a completely monotone function with a non-negative function that has a completely monotone derivative is again completely monotone. 
The densities for the outer power Clayton copula can be found in \cite{Ames2013}. 

In the above definitions of model based parametric upper and lower tail dependence one gets the estimates of joint extreme deviations in the whole currency basket. It will often be useful in practice to understand which pairs of currencies within a given currency basket contribute significantly to the downside or upside risks of the overall currency basket. In the class of Archimedean based mixtures we consider, the feature of exchangeability precludes decompositions of the total basket downside and upside risks into individual currency specific components. To be precise we aim to perform a decomposition of say the downside risk of the funding basket into contributions from each pair of currencies in the basket, we will do this is achieved via a simple linear projection onto particular subsets of currencies in the portfolio that are of interest, which leads for example to the following expression:
\begin{equation}
\mathbb{E}\left[\left.\hat  \lambda_u^{i|1,2,...,i-1,i+1,...,n} \right| \hat\lambda_u^{2|1}, \hat\lambda_u^{3|1}, \hat\lambda_u^{3|2},\ldots, \hat\lambda_u^{n|n-1}\right] = \alpha_0 + \sum_{i \neq j}^n \alpha_{ij}\hat\lambda_u^{i|j}, 
\end{equation}
where $\hat\lambda_u^{i|1,2,...,i-1,i+1,...,n}$ is a random variable since it is based on parameters of the mixture copula model which are themselves functions of the data and therefore random variables. Such a simple linear projection will then allow one to interpret directly the marginal linear contributions to the upside or downside risk exposure of the basket obtained from the model, according to particular pairs of currencies in the basket by considering the coefficients $\alpha_{ij}$, i.e. the projection weights. To perform this analysis we need estimates of the pairwise tail dependence in the upside and downside risk exposures $\hat\lambda_u^{i|j}$ and $\hat\lambda_l^{i|j}$ for each pair of currencies $i,j\in \left\{1,2,\ldots,n\right\}$. We obtain this through non-parametric (model-free) estimators, see \cite{Cruz2013}.

%

\begin{definition}Non-Parametric Pairwise Estimator of Upper Tail Dependence (Extreme Exposure) \\
\begin{equation}
\hat \lambda_u = 2 - \mbox{min} \left[2 \hspace{2mm}, \hspace{2mm} \frac{\mbox{log} \, \hat C_n \left( \frac{n - k}{n}, \frac{n - k}{n} \right)}{\mbox{log} (\frac{n - k}{n})} \right] \hspace{3mm} \hspace{3mm} k = {1,2, \ldots n-1},
\label{eq:nptd}
\end{equation}
where $\hat C_n \left( u_1, u_2 \right) = \frac{1}{n} \sum\limits_{i=1}^n \mathbf{1} \left( \frac{R_{1i}}{n} \leq u_1 , \frac{R_{2i}}{n} \leq u_2 \right)$
and $R_{ji}$ is the rank of the variable in its marginal dimension that makes up the pseudo data.
\end{definition}

In order to form a robust estimator of the upper tail dependence a median of the estimates obtained from setting $k$ as the $1^{st}, 2^{nd}, \ldots, 20^{th}$ percentile values was used. Similarly, $k$ was set to the $80^{th}, 81^{st}, \ldots, 99^{th}$ percentiles for the lower tail dependence.

%% file: FittingtheModels_GWP_26_Feb_2014_v1_0_0.tex


%
%

\section{Currency Basket Model Estimations via Inference Function For the Margins}
\label{section:likelihood}

The inference function for margins (IFM) technique introduced in \cite{Joe1996} provides a computationally faster method for estimating parameters than Full Maximum Likelihood, i.e. simultaneously maximising all model parameters and produces in many cases a more stable likelihood estimation procedure. 
This two stage estimation procedure was studied with regard to the asymptotic relative efficiency compared with maximum likelihood estimation in \cite{Joe2005} and in \cite{Hafner2010}. It can be shown that the IFM estimator is consistent under weak regularity conditions.



In modelling parametrically the marginal features of the log return forward exchange rates, we wanted flexibility to capture a broad range of skew-kurtosis relationships as well as potential for sub-exponential heavy tailed features. In addition, we wished to keep the models to a selection which is efficient to perform inference and easily interpretable. We consider a flexible three parameter model for the marginal distributions given by the Log-Generalized-Gamma distribution (l.g.g.d.), see details in \cite{lawless1980inference}, where
%
$Y$ has a l.g.g.d. if $Y = \mbox{log} (X)$ such that $X$ has a g.g.d. The density of $Y$ is given by
\begin{equation} \label{EqnLGGD}
f_{Y}(y; k,u,b) = \frac{1}{b \Gamma(k)}\exp\left[k\left(\frac{y - u}{b} \right) - \exp\left(\frac{y-u}{b}\right) \right],
\end{equation}
with $u = \mbox{log} \, (\alpha)$, $b = \beta^{-1}$ and the support of the l.g.g.d. distribution is $y \in \mathbb{R}$.

This flexible three parameter model admits
the LogNormal model as a limiting case (as $k \rightarrow \infty$). In addition the g.g.d. also includes the exponential model $(\beta=k=1)$, the Weibull distribution $(k=1)$ and the Gamma distribution $(\beta=1)$.


As an alternative to the l.g.g.d. model we also consider a time series approach to modelling the marginals,given by the GARCH($p$,$q$) model, as described in \cite{bollerslev1986generalized} and \cite{brechmann2012risk}, and characterised by the error variance:

\begin{equation}
\sigma^2 = \alpha_0 + \sum\limits_{i=1}^q \alpha_i \epsilon_{t-i}^2 + \sum\limits_{i=1}^p \beta_i \sigma_{t-i}^2 \;\; .
\end{equation}

\subsection{Stage 1: Fitting the Marginal Distributions via MLE}

The estimation for the three model parameters in the l.g.g.d. can be challenging due to the fact that a wide range of model parameters, especially for $k$, can produce similar resulting density shapes (see discussions in \cite{lawless1980inference}). To overcome this complication and to make the estimation efficient it is proposed to utilise a combination of profile likelihood methods over a grid of values for $k$ and perform profile likelihood based MLE estimation for each value of $k$, over the other two parameters $b$ and $u$. The differentiation of the profile likelihood for a given value of $k$ produces the system of two equations:
\begin{equation}
\exp(\tilde\mu) = \left[ \frac{1}{n}\sum_{i=1}^n \exp\left(\frac{y_i}{\tilde\sigma\sqrt{k}}\right) \right]^{\tilde\sigma \sqrt{k}}\\
\hspace{5mm} ; \hspace{5mm}
\frac{\sum_{i=1}^n y_i \exp\left(\frac{y_i}{\tilde\sigma\sqrt{k}}\right)}{\sum_{i=1}^n \exp\left(\frac{y_i}{\tilde\sigma\sqrt{k}}\right)} - \overline{y} - \frac{\tilde\sigma}{\sqrt{k}} = 0 \; ,
\label{lggd_mle}
\end{equation}
where $n$ is the number of observations, $y_i = \mbox{log} \, x_i$, $\tilde\sigma = b/\sqrt{k}$ and $\tilde\mu = u + b \, \mbox{log} \,k$. The second equation is solved directly via a simple root search to give an estimation for $\tilde{\sigma}$ and then substitution into the first equation results in an estimate for $\tilde{\mu}$. Note, for each value of $k$ we select in the grid, we get the pair of parameter estimates $\tilde\mu$ and $\tilde\sigma$, which can then be plugged back into the profile likelihood to make it purely a function of $k$, with the estimator for $k$ then selected as the one with the maximum likelihood score.
As a comparison we also fit the GARCH(1,1) model using the MATLAB MFEtoolbox using the default settings.

\subsection{Stage 2: Fitting the Mixture Copula via MLE}

In order to fit the copula model the parameters are estimated using maximum likelihood on the data after conditioning on the selected marginal distribution models and their corresponding estimated parameters obtained in Stage 1. These models are utilised to transform the data using the CDF function with the l.g.g.d. MLE parameters ($\hat k$, $\hat u$ and $\hat b$) or using the conditional variances to obtain standardised residuals for the GARCH model. Therefore, in this second stage of MLE estimation we aim to estimate either the one parameter mixture of CFG components with parameters ${\bf\underline\theta} = (\rho_{clayton}, \rho_{frank}, \rho_{gumbel}, \lambda_{clayton}, \lambda_{frank}, \lambda_{gumbel})$, the one parameter mixture of CG components with parameters ${\bf\underline\theta} = (\rho_{clayton}, \rho_{gumbel}, \lambda_{clayton}, \lambda_{gumbel})$  or the two parameter outer power transformed Clayton with parameters ${\bf\underline\theta} = (\rho_{clayton}, \beta_{clayton})$. 
The log likelihood expression for the mixture copula models, is given generically by:
\begin{equation}
l({\bf\underline\theta}) = \sum_{i=1}^n \mbox{log} \; c(F_1(X_{i1};\hat\mu_1, \hat\sigma_1), \dots, F_d(X_{id};\hat\mu_d, \hat\sigma_d))  \; + \; \sum_{i=1}^n \sum_{j=1}^d \mbox{log}  \;f_j(X_{ij};\hat\mu_j, \hat\sigma_j).
\label{loglik}
\end{equation}


This optimization is achieved via a gradient descent iterative algorithm which was found to be quite robust given the likelihood surfaces considered in these models with the real data. Alternative estimation procedures such as expectation-maximisation were not found to be required.






\FloatBarrier

%% file: DataDescriptionBasketFormation_GWP_27_Feb_2014_v1_0_0.tex
\section{Exchange Rate Multivariate Data Description and Currency Portfolio Construction}

In our study we fit copula models to the high interest rate basket and the low interest rate basket updated for each day in the period 02/01/1989 to 29/01/2014 using log return forward exchange rates at one month maturities for data covering both the previous 6 months and previous year as a sliding window analysis on each trading day in this period. 


Our empirical analysis consists of daily exchange rate data for a set of 34 currency exchange rates relative to the USD, as in \cite{Menkhoff2012}. The currencies analysed included: Australia (AUD), Brazil (BRL), Canada (CAD), Croatia (HRK), Cyprus (CYP), Czech Republic (CZK), Egypt (EGP), Euro area (EUR), Greece (GRD), Hungary (HUF), Iceland (ISK), India (INR), Indonesia (IDR), Israel (ILS), Japan (JPY), Malaysia (MYR), Mexico (MXN), New Zealand (NZD), Norway (NOK), Philippines (PHP), Poland (PLN), Russia (RUB), Singapore (SGD), Slovakia (SKK), Slovenia (SIT), South Africa (ZAR), South Korea (KRW), Sweden (SEK), Switzerland (CHF), Taiwan (TWD), Thailand (THB), Turkey (TRY), Ukraine (UAH) and the United Kingdom (GBP).

We have considered daily settlement prices for each currency exchange rate as well as the daily settlement price for the associated 1 month forward contract. We utilise the same dataset (albeit starting in 1989 rather than 1983 and running up until January 2014) as studied in \cite{Lustig2011} and \cite{Menkhoff2012} in order to replicate their portfolio returns without tail dependence risk adjustments. Due to differing market closing days, e.g. national holidays, there was missing data for a couple of currencies and for a small number of days. For missing prices, the previous day's closing prices were retained. 

As was demonstrated in Equation (\ref{UIP}), the differential of interest rates between two countries can be estimated through the ratio of the forward contract price and the spot price, see \cite{Juhl2006} who show this holds empirically on a daily basis. Accordingly, instead of considering the differential of risk free rates between the reference and the foreign countries, we build our respective baskets of currencies with respect to the ratio of the forward and the spot prices for each currency. On a daily basis we compute this ratio for each of the $n$ currencies (available in the dataset on that day) and then build five baskets. The first basket gathers the $n/5$ currencies with the highest positive differential of interest rate with the US dollar.
These currencies are thus representing the ``investment" currencies, through which we invest the money to benefit from the currency carry trade. The last basket will gather the $n/5$ currencies with the highest negative differential (or at least the lowest differential) of interest rate.
These currencies are thus representing the ``financing" currencies, through which we borrow the money to build the currency carry trade.

Given this classification we investigate then the joint distribution of each group of currencies to understand the impact of the currency carry trade, embodied by the differential of interest rates, on currencies returns. In our analysis we concentrate on the high interest rate basket (investment currencies) and the low interest rate basket (funding currencies), since typically when implementing a carry trade strategy one would go short the low interest rate basket and go long the high interest rate basket.

%% file: Results_GWP_27_Feb_2014_v1.tex
\section{Results and Discussion}


In order to model the marginal exchange rate log-returns we considered two approaches. Firstly, we fit Log Generalised Gamma models to each of the 34 currencies considered in the analysis, updating the fits for every trading day based on a 6 month sliding window. A time series approach was also considered to fit the marginals, as is popular in much of the recent copula literature, see for example \cite{brechmann2012risk}, using GARCH(1,1) models for the 6 month sliding data windows. In each case we are assuming approximate local stationarity over these short 6 month time frames.

A summary of the marginal model selection can be seen in Table~\ref{AIC_margins}, which shows the average AIC scores for the 4 most frequent currencies in the high interest rate and the low interest rate baskets over the data period. Whilst the AIC for the GARCH(1,1) model is consistently lower than the respective AIC for the Generalised Gamma, the standard errors are sufficiently large for there to be no clear favourite between the two models.

\renewcommand{\arraystretch}{1.5}
\setlength{\tabcolsep}{8pt}

\begin{table}
  \centering
  \caption{Average AIC for the Generalized Gamma (GG) and the GARCH(1,1) for the four most frequent currencies in the high interest rate and the low interest rate baskets over the 2001 - 2014 data period split into 2 chunks, i.e. 6 years. Standard deviations are shown in parentheses. Similar performance was seen between 1989-2001.}
    \begin{tabular}{cccccc}
           \toprule
          & \textbf{} & \multicolumn{2}{c}{\textbf{01 - 07}} & \multicolumn{2}{c}{\textbf{07 - 14}} \\
           \midrule
          & \textbf{Currency} & \textbf{GG} & \textbf{GARCH} & \textbf{GG} & \textbf{GARCH} \\
    \multirow{4}[0]{*}{\begin{sideways}Investment\end{sideways}} & \textbf{TRY} & 356.9 (3.5) & \textbf{341.1 (21.7)} & 358.7 (3.0) & \textbf{349.1 (16.8)} \\
          & \textbf{MXN} & 360.0 (1.2) & \textbf{357.04 (3.8)} & 358.6 (4.0) & \textbf{344.5 (28.1)} \\
          & \textbf{ZAR} & 358.7 (3.0) & \textbf{353.5 (11.4)} & 358.0 (6.1) & \textbf{352.8 (12.2)} \\
          & \textbf{BRL} & 359.0 (2.8) & \textbf{341.6 (19.4)} & 360.0 (2.1) & \textbf{341.6 (23.2)} \\
          \midrule	
    \multirow{4}[0]{*}{\begin{sideways}Funding\end{sideways}} & \textbf{JPY} & 361.2 (0.9) & \textbf{356.5 (7.2)} & 356.9 (6.8) & \textbf{355.0 (7.0)} \\
          & \textbf{CHF} & 360.8 (1.4) & \textbf{359.1 (2.9)} & 358.6 (7.4) & \textbf{355.4 (8.8)} \\
          & \textbf{SGD} & 360.0 (2.7) & \textbf{356.8 (5.7)} & 360.0 (2.6) & \textbf{353.7 (7.5)} \\
          & \textbf{TWD} & 358.7 (6.2) & \textbf{347.0 (16.4)} & 359.1 (5.8) & \textbf{348.5 (13.2)} \\
    \bottomrule
    \end{tabular}%
  \label{AIC_margins}%
\end{table}%

\begin{figure}
\centering
\includegraphics[width =\textwidth , height = 60mm]{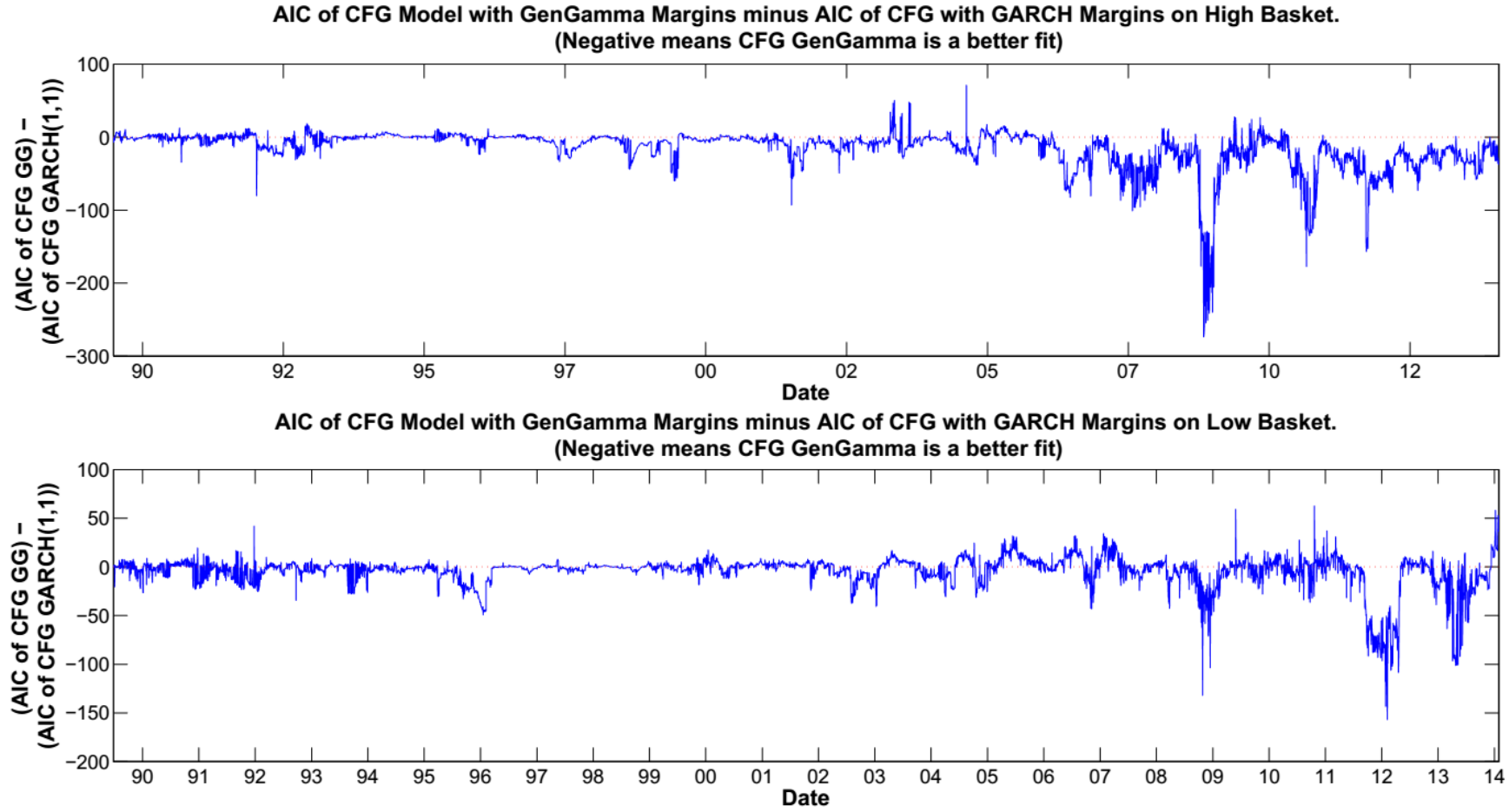}
\caption{Comparison of AIC for Clayton-Frank-Gumbel model fit on the pseudo data resulting from Generalised Gamma vs GARCH(1,1) margins . The high interest rate basket is shown in the upper panel and the low interest rate basket is shown in the lower panel.}
\label{CFG_GG_vs_GARCH_high_and_low}
\end{figure}

However, when we consider the model selection of the copula in combination with the marginal model we observe lower AIC  scores for copula models fitted on the pseudo data resulting from using Generalised Gamma margins than using GARCH(1,1) margins. This is the case for all three copula models under consideration in the paper. Figure~\ref{CFG_GG_vs_GARCH_high_and_low} shows the AIC differences when using the Clayton-Frank-Gumbel copula in combination with the two choices of marginal for the high interest rate and the low interest rate basket respectively.
Over the entire data period the mean difference between the AIC scores for the CFG model with Generalised Gamma vs GARCH(1,1) marginals for the high interest rate basket is 12.3 and for the low interest rate basket is 3.6 in favour of the Generalised Gamma.

Thus, it is clear that the Generalised Gamma model is the better model in our copula modelling context and so is used in the remainder of the analysis. We now consider the goodness-of-fit of the three copula models applied to the high interest rate basket and low interest rate basket pseudo data. We used a scoring via the AIC between the three component mixture CFG model versus the two component mixture CG model versus the two parameter OpC model. One could also use the Copula-Information-Criterion (CIC), see \cite{Gronneberg2010} for details. 

\begin{figure}
\centering
\includegraphics[width =\textwidth , height = 60mm]{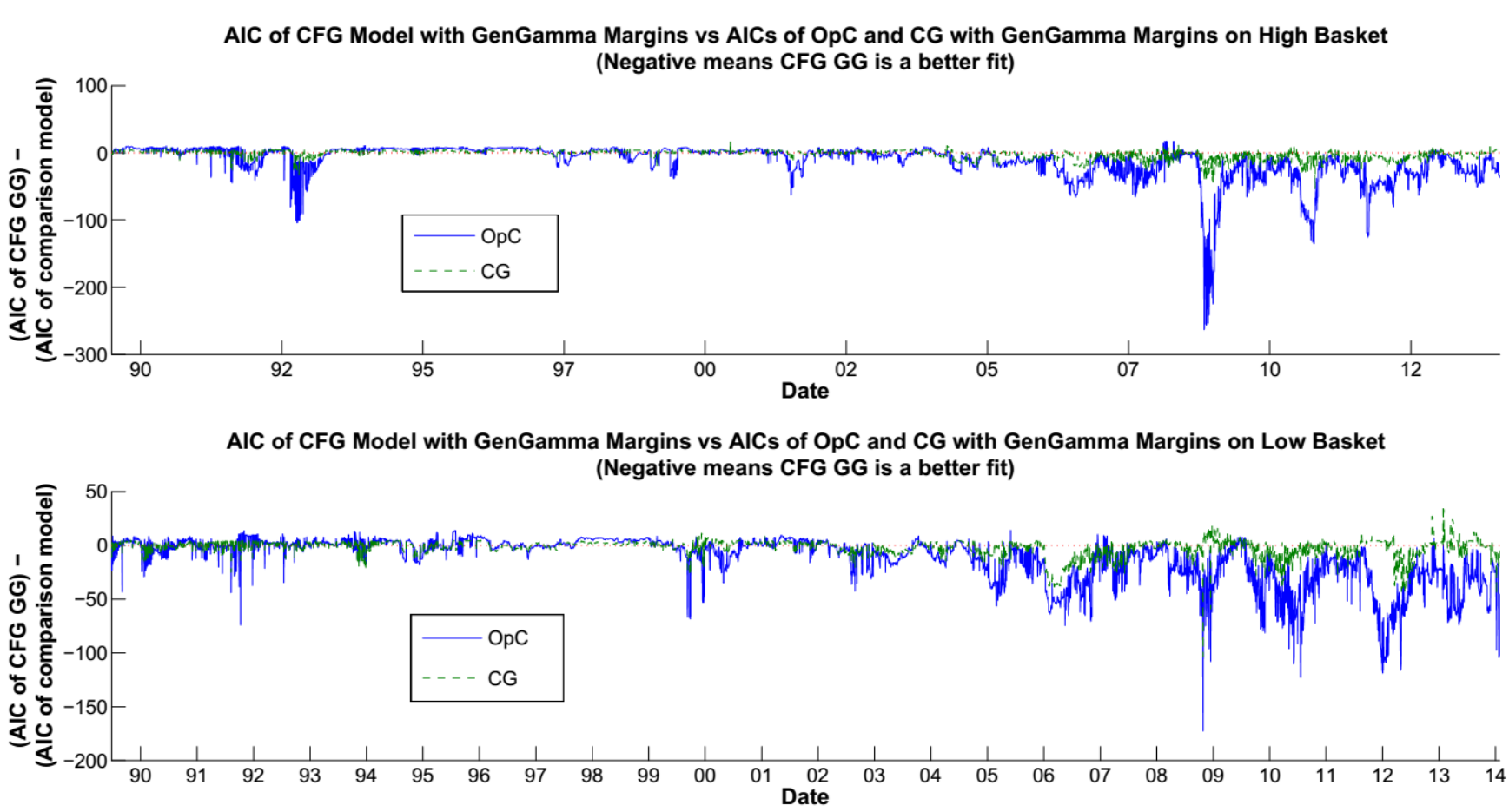}
\caption{Comparison of AIC for Clayton-Frank-Gumbel model with Clayton-Gumbel and Outer power Clayton models on high and low interest rate baskets with Generalised Gamma margins. The high interest rate basket is shown in the upper panel and the low interest rate basket is shown in the lower panel.}
\label{CFG_GG_vs_CG_OpC_high_and_low}
\end{figure}

The results are presented for this comparison in Figure~\ref{CFG_GG_vs_CG_OpC_high_and_low}, which shows the differentials between AIC for CFG versus CG and CFG versus OpC for each of the high interest rate and the low interest rate currency baskets. We can see it is not unreasonable to consider the CFG model for this analysis, since over the entire data period the mean difference between the AIC scores for the CFG and the CG models for the high interest rate basket is 1.33 and for the low interest rate basket is 1.62 in favour of the CFG.

However, from Figure~\ref{CFG_GG_vs_CG_OpC_high_and_low} we can see that during the Credit crisis period the CFG model is performing much better.
The CFG copula model provides a much better fit when compared to the OpC model, as shown by the mean difference between the AIC scores of 9.58 for the high interest rate basket and 9.53 for the low interest rate basket. Similarly, the CFG model performs markedly better than the OpC model during the Credit crisis period.

\subsection{Tail Dependence Results}
\begin{figure}
\centering
\includegraphics[width =\textwidth , height=80mm]{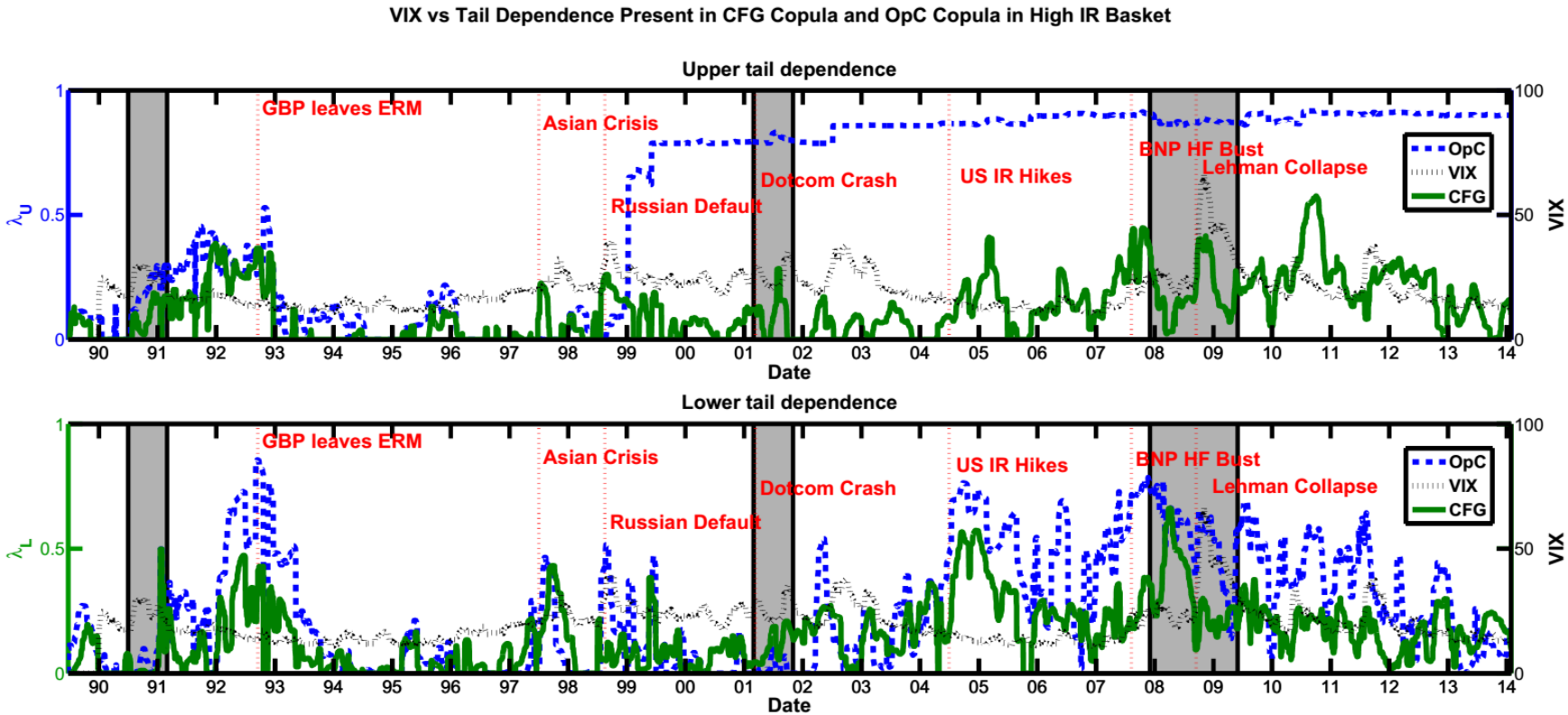}
\caption{Comparison of Volatility Index (VIX) with upper and lower tail dependence of the high interest rate basket in the CFG copula and OpC copula. US NBER recession periods are represented by the shaded grey zones. Some key crisis dates across the time period are labelled.}
\label{CFG_OPC_VIX_high_TD}
\end{figure}

\begin{figure}
\centering
\includegraphics[width =\textwidth , height=80mm]{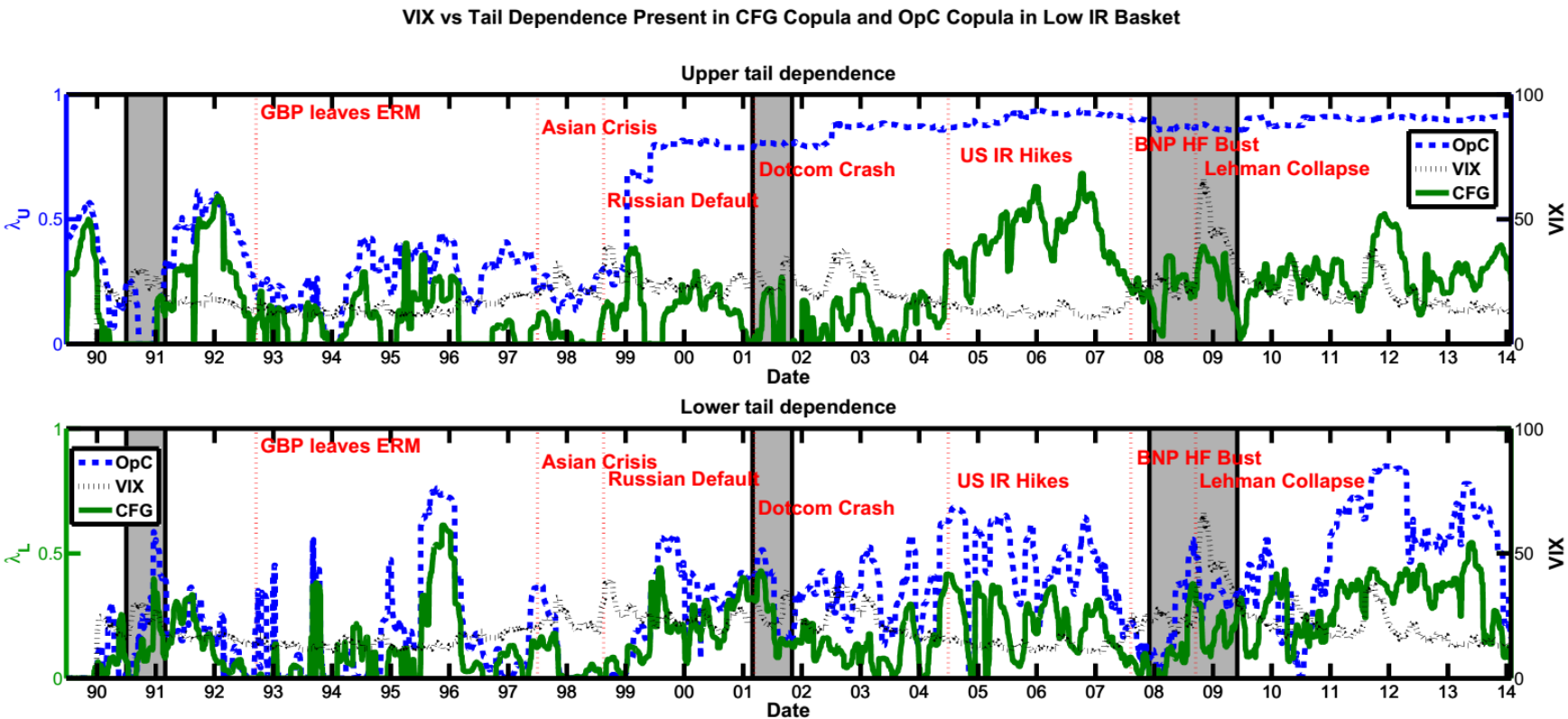}
\caption{Comparison of Volatility Index (VIX) with upper and lower tail dependence of the low interest rate basket in the CFG copula and OpC copula. US NBER recession periods are represented by the shaded grey zones. Some key crisis dates across the time period are labelled.}
\label{CFG_OPC_VIX_low_TD}
\end{figure}

Below we will examine the time-varying parameters of the maximum likelihood fits of this mixture CFG copula model. Here, we shall focus on the strength of dependence present in the currency baskets, given the particular copula structures in the mixture, which is considered as tail upside/downside exposure of a carry trade over time. Figure~\ref{CFG_OPC_VIX_high_TD} shows the time-varying upper and lower tail dependence, i.e. the extreme upside and downside risk exposures for the carry trade basket, present in the high interest rate basket under the CFG copula fit and the OpC copula fit. Similarly, Figure~\ref{CFG_OPC_VIX_low_TD} shows this for the low interest rate basket. 

\begin{remark}[Model Risk and its Influence on Upside and Downside Risk Exposure]
In fitting the OpC model, we note that independent of the strength of true tail dependence in the multivariate distribution, the upper tail dependence coefficient $\lambda_u$ for this model strictly increases with dimension very rapidly. Therefore, when fitting the OpC model, if the basket size becomes greater than bivariate, i.e. from 1999 onwards, the upper tail dependence estimates become very large (even for outer-power parameter values very close to $\beta=1$). This lack of flexibility in the OpC model only becomes apparent in baskets of dimension greater than 2, but is also evident in the AIC scores in Figure~\ref{CFG_GG_vs_CG_OpC_high_and_low}. Here we see an interesting interplay between the model risk associated to the dependence structure being fit and the resulting interpreted upside or downside financial risk exposures for the currency baskets.
\end{remark}

Focusing on the tail dependence estimate produced from the CFG copula fits we can see that there are indeed periods of heightened upper and lower tail dependence in the high interest rate and the low interest rate baskets. There is a noticeable increase in upper tail dependence in the high interest rate basket at times of global market volatility. Specifically, during late 2007, i.e. the global financial crisis, there is a sharp peak in upper tail dependence. Preceding this, there is an extended period of heightened lower tail dependence from 2004 to 2007, which could tie in with the building of the leveraged carry trade portfolio positions. This period of carry trade construction is also very noticeable in the low interest rate basket through the very high levels of upper tail dependence.

We compare in Figures~\ref{CFG_OPC_VIX_high_TD} and \ref{CFG_OPC_VIX_low_TD} the tail dependence plotted against the VIX volatility index for the high interest rate basket and the low interest rate basket respectively for the period under investigation. The VIX is a popular measure of the implied volatility of S\&P 500 index options - often referred to as the \emph{fear index}. As such it is one measure of the market's expectations of stock market volatility over the next 30 days. We can clearly see here that in the high interest rate basket there are upper tail dependence peaks at times when there is an elevated VIX index, particularly post-crisis. However, we would not expect the two to match exactly since the VIX is not a direct measure of global FX volatility. We can thus conclude that investors' risk aversion clearly plays an important role in the tail behaviour. This conclusion corroborates recent literature regarding the skewness and the kurtosis features characterizing the currency carry trade portfolios \cite{Farhi2008,Brunnermeier2008, Menkhoff2012}.

\subsection{Pairwise Decomposition of Basket Tail Dependence}

\begin{figure}
\centering
\includegraphics[width =\textwidth, height=75mm]{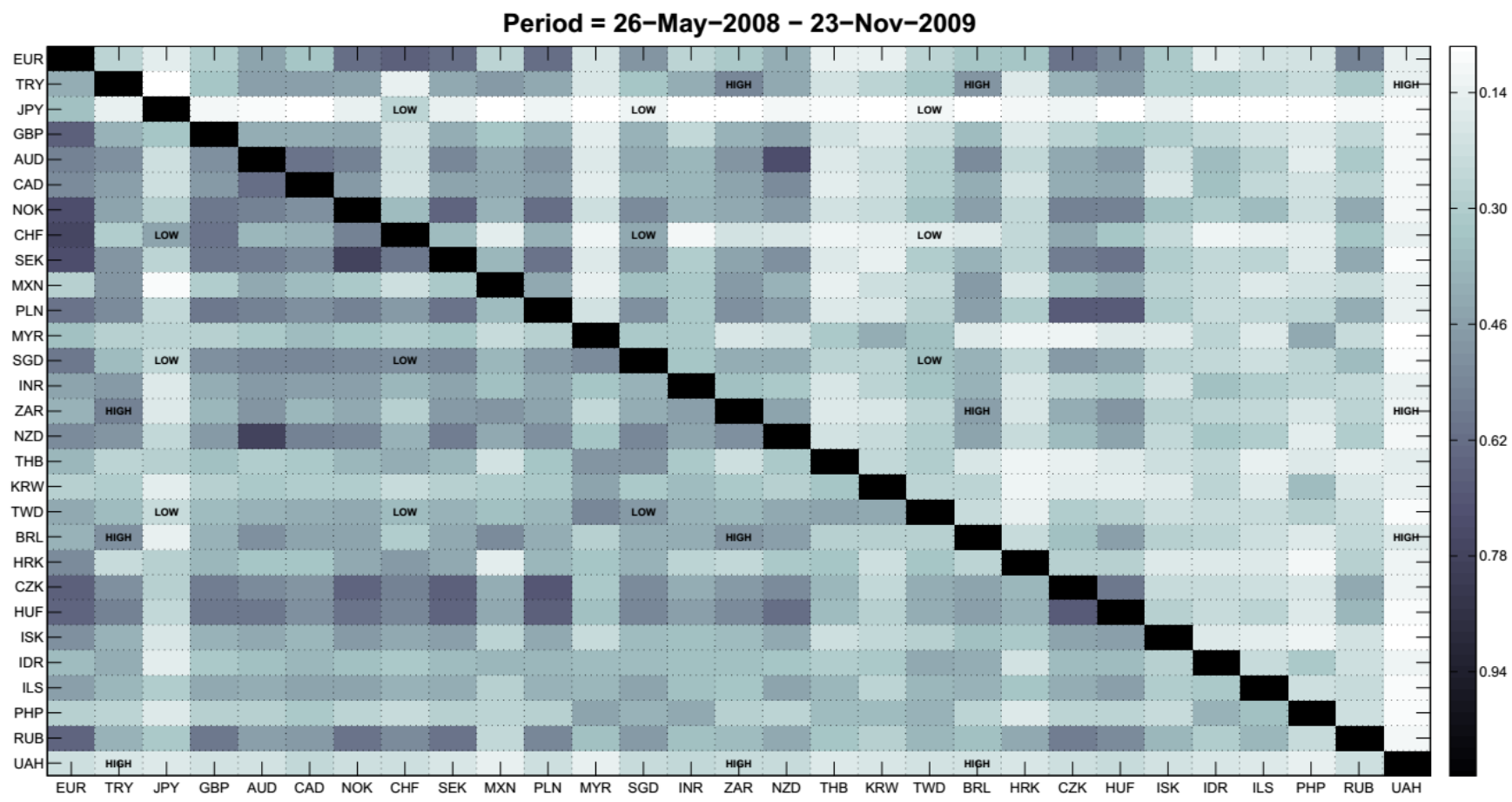}	%
\caption{Heat map showing the strength of non-parametric tail dependence between each pair of currencies averaged over the Credit crisis period. Lower tail dependence is shown in the lower triangle and upper tail dependence is shown in the upper triangle. The 3 currencies most frequently in the high interest rate and the low interest rate baskets are labelled.}
\label{crisis3_heatmap}
\end{figure}

In order to examine the contribution of each pair of currencies to the overall n-dimensional basket tail dependence we calculated the corresponding non-parametric pairwise tail dependencies for each pair of currencies. In Figure~\ref{crisis3_heatmap} we can see the average upper and lower non-parametric tail dependence for each pair of currencies during the Credit crisis, with the 3 currencies most frequently in the high interest rate and the low interest rate baskets labelled accordingly. The lower triangle represents the non-parametric pairwise lower tail dependence and the upper triangle represents the non-parametric pairwise upper tail dependence.

If one was trying to optimise their currency portfolio with respect to the tail risk exposures, i.e. to minimise negative tail risk exposure and maximise positive tail risk exposure, then one would sell short currencies with high upper tail dependence and low lower tail dependence whilst buying currencies with low upper tail dependence and high lower tail dependence.

\begin{figure}
\centering
\includegraphics[width =\textwidth, height=75mm]{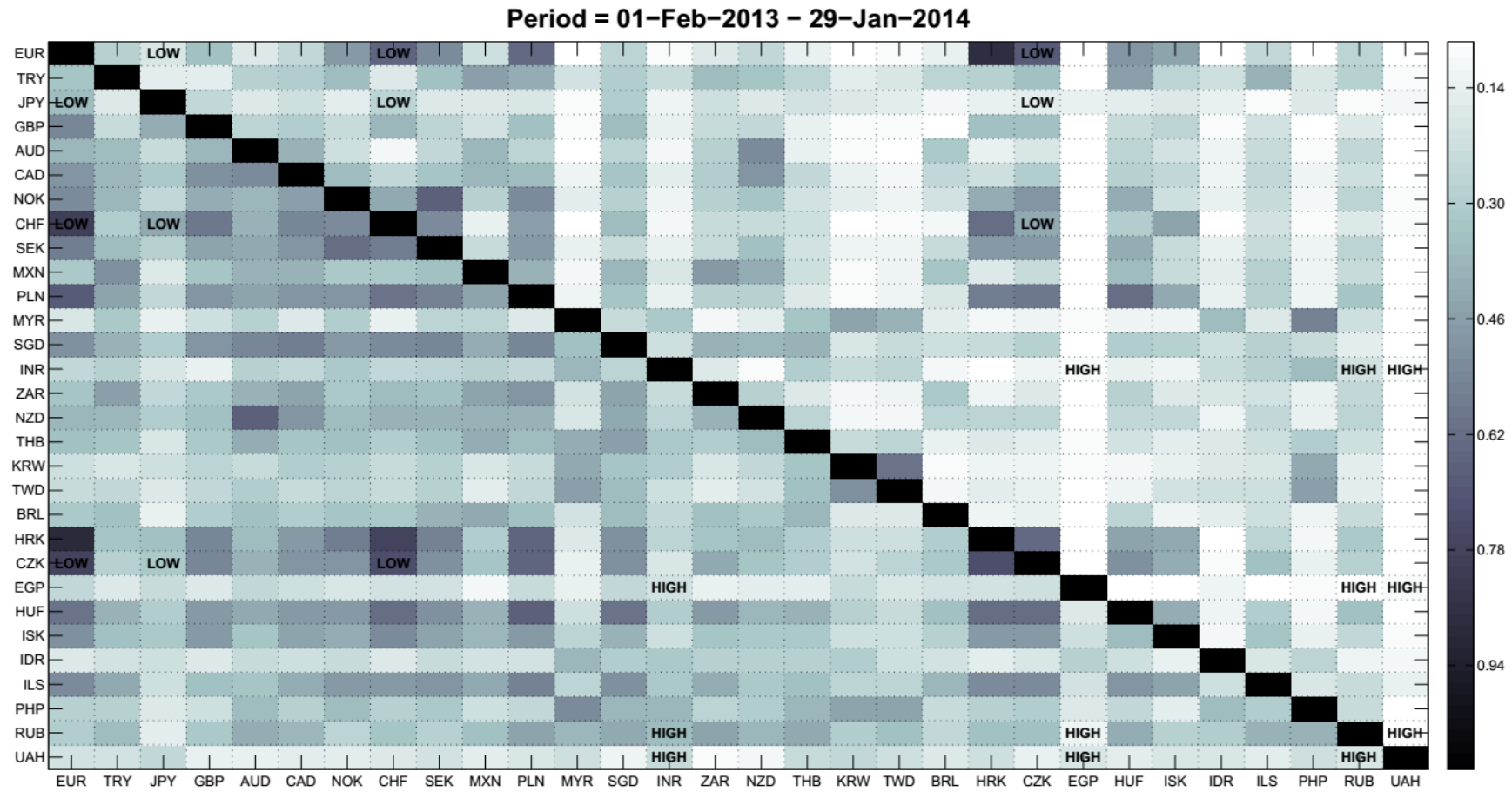}
\caption{Heat map showing the strength of non-parametric tail dependence between each pair of currencies averaged over the last 12 months (01/02/2013 to 29/01/2014). Lower tail dependence is shown in the lower triangle and upper tail dependence is shown in the upper triangle. The 3 currencies most frequently in the high interest rate and the low interest rate baskets are labelled.}
\label{last_12_months_heatmap}
\end{figure}

Similarly, in Figure~\ref{last_12_months_heatmap} we see the pairwise non-parametric tail dependencies averaged over the last 12 months (01/02/2013 to 29/01/2014). Comparing this heat map to the heat map during the Credit crisis (Figure~\ref{crisis3_heatmap}) we notice that in general there are lower values of tail dependence amongst the currency pairs.


We performed linear regression of the pairwise non-parametric tail dependence on the respective basket tail dependence for the days on which the 3 currencies all appeared in the basket (224 out of 250 for the lower interest rate basket and 223 out of 250 for the high interest rate basket). The regression coefficients and $R^2$ values can be seen in Table~\ref{nptd_td_regression}. We can interpret this as the relative contribution of each of the 3 currency pairs to the overall basket tail dependence. We note that for the low interest rate lower tail dependence and for the high interest rate upper tail dependence there is a significant degree of cointegration between the currency pair covariates and hence we might be able to use a single covariate due to the presence of a common stochastic trend.


\setlength{\tabcolsep}{6pt}
\begin{table}
  \centering
  \caption{Pairwise non-parametric tail dependence regressed on respective basket tail dependence (standard errors are shown in parentheses). The 3 currencies most frequently in the respective baskets are used as independent variables.}
    \begin{tabular}{cccccc}
    \toprule
    \textbf{Low IR Basket } & \textbf{Constant} & \textbf{CHF JPY} & \textbf{CZK CHF} & \textbf{CZK JPY} & \textbf{$R^2$} \\
    \midrule
    \textbf{Upper TD} & 0.22 (0.01)  & 0.02 (0.03)  & 0.18 (0.02)  & 0.38 (0.05)  & 0.57 \\
    \textbf{Lower TD} & 0.71 (0.17)  & -0.62 (0.25) & -0.38 (0.26) & 0.23 (0.32)  & 0.28 \\
    \bottomrule
    \textbf{} &       &       &       &       &  \\
        \toprule
    \textbf{High IR Basket } & \textbf{Constant} & \textbf{EGP INR} & \textbf{UAH EGP} & \textbf{UAH INR} & \textbf{$R^2$} \\
        \midrule
    \textbf{Upper TD} & 0.07 (0.01)  & -0.06 (0.33) & 0.59 (0.08)  & 2.37 (0.42)  & 0.4 \\
    \textbf{Lower TD} & 0.1 (0.02)   & 0.56 (0.05)  & 0.44 (0.08)  & -0.4 (0.07)  & 0.44 \\
    \bottomrule
    \end{tabular}%
  \label{nptd_td_regression}%
\end{table}%

\subsection{Understanding the Tail Exposure associated with the Carry Trade and its Role in the UIP Puzzle}


As was discussed in Section~\ref{joint_tail_risk_exposure}, the tail exposures associated with a currency carry trade strategy can be broken down into the upside and downside tail exposures within each of the long and short carry trade baskets.
The downside relative exposure adjusted returns are obtained by multiplying the monthly portfolio returns by one minus the upper and the lower tail dependence present respectively in the high interest rate basket and the low interest rate basket at the corresponding dates. The upside relative exposure adjusted returns are obtained by multiplying the monthly portfolio returns by one plus the lower and upper tail dependence present respectively in the high interest rate basket and the low interest rate basket at the corresponding dates. Note that we refer to these as relative exposure adjustments only for the tail exposures since we do not quantify a market price per unit of tail risk. However, this is still informative as it shows a decomposition of the relative exposures from the long and short baskets with regard to extreme events.

\begin{figure}
\centering
\includegraphics[width =\textwidth , height=75mm]{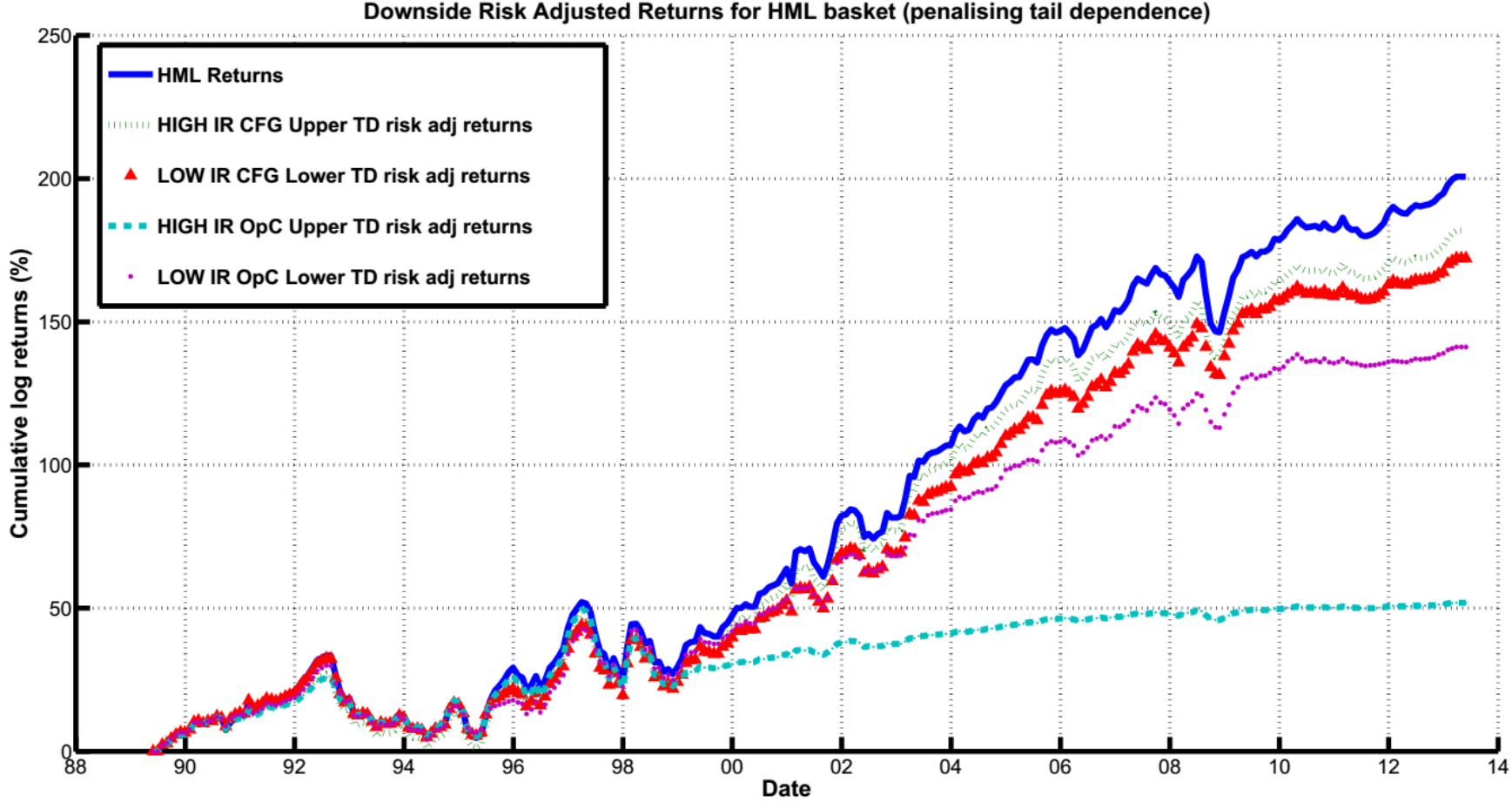}
\caption{Cumulative log returns of the carry trade portfolio (HML = High interest rate basket Minus Low interest rate basket). Downside exposure adjusted cumulative log returns using upper/lower tail dependence in the high/low interest rate basket for the CFG copula and the OpC copula are shown for comparison.}
\label{risk_adj_Downside}
\end{figure}

\begin{figure}
\centering
\includegraphics[width =\textwidth ,height=75mm]{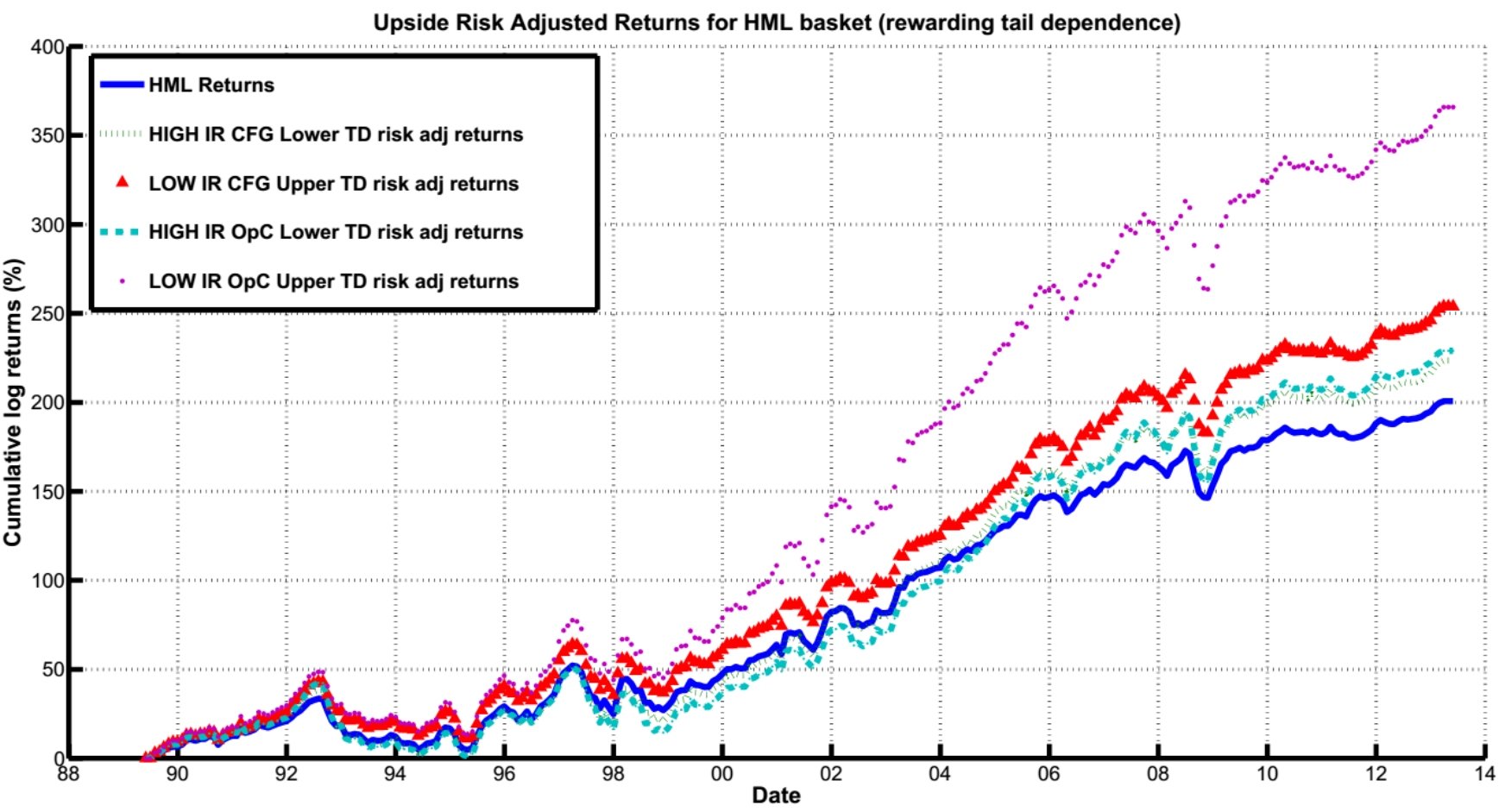}
\caption{Cumulative log returns of the carry trade portfolio (HML = High interest rate basket Minus Low interest rate basket). Upside exposure adjusted cumulative log returns using lower/upper tail dependence in the high/low interest rate basket for the CFG copula and the OpC copula are shown for comparison.}
\label{risk_adj_Upside}
\end{figure}

As can be seen in Figure~\ref{risk_adj_Downside}, the relative adjustment to the absolute cumulative returns for each type of downside exposure is greatest for the low interest rate basket, except under the OpC model, but this is due to the very poor fit of this model to baskets containing more than 2 currencies which we see transfers to financial risk exposures. This is interesting because intuitively one would expect the high interest rate basket to be the largest source of tail exposure. However, one should be careful when interpreting this plot, since we are looking at the extremal tail exposure. The analysis may change if one considered the intermediate tail risk exposure, where the marginal effects become significant.
Similarly, Figure~\ref{risk_adj_Upside} shows the relative adjustment to the absolute cumulative returns for each type of upside exposure is greatest for the low interest rate basket. The same interpretation as for the downside relative exposure adjustments can be made here for upside relative exposure adjustments.



%% file: Discussion_MA_14_Nov_2013_v3_0_0.tex
\section{Conclusion}
\label{conclusion}

In this paper, we have shown that the positive and negative multivariate tail risk exposures present in currency carry trade baskets are additional factors needing careful consideration when one constructs a carry portfolio. Ignoring these exposures leads to a perceived risk return profile that is not reflective of the true nature of such a strategy.
In terms of marginal model selection, it was shown that one is indifferent between the log Generalised Gamma model and the frequently used GARCH(1,1) model. However, in combination with the three different Archimedean copula models considered in this paper the log Generalised Gamma marginals provided a better overall model fit.

%% file: _Matthew_Ames__Innovations_in_Risk_Management__Revised_.bbl
\begin{thebibliography}{10}

\bibitem{Ames2013}
M.~Ames, G.~Bagnarosa, and G.~W. Peters.
\newblock {R}einvestigating the {U}ncovered {I}nterest {R}ate {P}arity {P}uzzle
  via {A}nalysis of {M}ultivariate {T}ail {D}ependence in {C}urrency {C}arry
  {T}rades.
\newblock {\em arXiv:1303.4314}, 2013.

\bibitem{backus2001affine}
D.~K. Backus, S.~Foresi, and C.~I. Telmer.
\newblock {A}ffine {T}erm {S}tructure {M}odels and the {F}orward {P}remium
  {A}nomaly.
\newblock {\em The Journal of Finance}, 56(1):279--304, 2001.

\bibitem{bollerslev1986generalized}
T.~Bollerslev.
\newblock {G}eneralized {A}utoregressive {C}onditional {H}eteroskedasticity.
\newblock {\em Journal of Econometrics}, 31(3):307--327, 1986.

\bibitem{brechmann2012risk}
E.~C. Brechmann and C.~Czado.
\newblock {R}isk {M}anagement with {H}igh-dimensional {V}ine {C}opulas: {A}n
  {A}nalysis of the {E}uro {S}toxx 50.
\newblock {\em Statistics \& Risk Modeling,}, 30(4):307--342, 2012.

\bibitem{Brunnermeier2008}
M.~K. Brunnermeier, S.~Nagel, and L.~H. Pedersen.
\newblock {C}arry {T}rades and {C}urrency {C}rashes.
\newblock Working Paper 14473, National Bureau of Economic Research, November
  2008.

\bibitem{burnside2011peso}
C.~Burnside, M.~Eichenbaum, I.~Kleshchelski, and S.~Rebelo.
\newblock {D}o {P}eso {P}roblems {E}xplain the {R}eturns to the {C}arry
  {T}rade?
\newblock {\em Review of Financial Studies}, 24(3):853--891, 2011.

\bibitem{christiansen2011time}
C.~Christiansen, A.~Ranaldo, and P.~S{\"o}derlind.
\newblock {T}he {T}ime-varying {S}ystematic {R}isk of {C}arry {T}rade
  {S}trategies.
\newblock {\em Journal of Financial and Quantitative Analysis},
  46(04):1107--1125, 2011.

\bibitem{Cruz2013}
M.~Cruz, G.~Peters, and P.~Shevchenko.
\newblock {\em {H}andbook on {O}perational {R}isk}.
\newblock Wiley New York, 2013.

\bibitem{de2012multivariate}
G.~De~Luca and G.~Rivieccio.
\newblock {M}ultivariate {T}ail {D}ependence {C}oefficients for {A}rchimedean
  {C}opulae.
\newblock {\em Advanced Statistical Methods for the Analysis of Large
  Data-Sets}, page 287, 2012.

\bibitem{Fama1984}
E.~F. Fama.
\newblock {F}orward and {S}pot {E}xchange {R}ates.
\newblock {\em Journal of Monetary Economics}, 14(3):319--338, 1984.

\bibitem{Farhi2008}
E.~Farhi and X.~Gabaix.
\newblock {R}are {D}isasters and {E}xchange {R}ates.
\newblock Working Paper 13805, National Bureau of Economic Research, February
  2008.

\bibitem{feller1971}
W.~Feller.
\newblock {\em {A}n {I}ntroduction to {P}robability {T}heory and {I}ts
  {A}pplications. {V}ol. 2}.
\newblock New York: Wiley, 1971.

\bibitem{Gronneberg2010}
S.~Gr{\o}nneberg.
\newblock {T}he {C}opula {I}nformation {C}riterion and its {I}mplications for
  the {M}aximum {P}seudo-likelihood {E}stimator.
\newblock {\em Dependence Modelling: The Vine Copula Handbook, World Scientific
  Books}, pages 113--138, 2010.

\bibitem{Hafner2010}
C.~M. Hafner and H.~Manner.
\newblock {D}ynamic {S}tochastic {C}opula {M}odels: {E}stimation, {I}nference
  and {A}pplications.
\newblock {\em Journal of Applied Econometrics}, 27(2):269--295, 2010.

\bibitem{Hansen1980}
L.~P. Hansen and R.~J. Hodrick.
\newblock {F}orward {E}xchange {R}ates as {O}ptimal {P}redictors of {F}uture
  {S}pot {R}ates: {A}n {E}conometric {A}nalysis.
\newblock {\em The Journal of Political Economy}, pages 829--853, 1980.

\bibitem{Joe2005}
H.~Joe.
\newblock {A}symptotic {E}fficiency of the {T}wo-stage {E}stimation {M}ethod
  for {C}opula-based {M}odels.
\newblock {\em Journal of Multivariate Analysis}, 94(2):401--419, 2005.

\bibitem{Joe1996}
H.~Joe and J.~J. Xu.
\newblock {T}he {E}stimation {M}ethod of {I}nference {F}unctions for {M}argins
  for {M}ultivariate {M}odels.
\newblock Technical report, Technical Report 166, Department of Statistics,
  University of British Columbia, 1996.

\bibitem{Juhl2006}
T.~Juhl, W.~Miles, and M.~D. Weidenmier.
\newblock {C}overed {I}nterest {A}rbitrage: {T}hen versus {N}ow.
\newblock {\em Economica}, 73(290):341--352, 2006.

\bibitem{lawless1980inference}
J.~F. Lawless.
\newblock {I}nference in the {G}eneralized {G}amma and {L}og {G}amma
  {D}istributions.
\newblock {\em Technometrics}, 22(3):409--419, 1980.

\bibitem{Lustig2011}
H.~Lustig, N.~Roussanov, and A.~Verdelhan.
\newblock {C}ommon {R}isk {F}actors in {C}urrency {M}arkets.
\newblock {\em Review of Financial Studies}, 24(11):3731--3777, 2011.

\bibitem{Lustig2007}
H.~Lustig and A.~Verdelhan.
\newblock {T}he {C}ross {S}ection of {F}oreign {C}urrency {R}isk {P}remia and
  {C}onsumption {G}rowth {R}isk.
\newblock {\em American Economic Review}, 97(1):89--117, September 2007.

\bibitem{McNeil2009}
A.~J. McNeil and J.~Ne{\v{s}}lehov{\'a}.
\newblock {M}ultivariate {A}rchimedean {C}opulas, d-monotone {F}unctions and
  {L}1-norm {S}ymmetric {D}istributions.
\newblock {\em The Annals of Statistics}, 37(5B):3059--3097, 2009.

\bibitem{Menkhoff2012}
L.~Menkhoff, L.~Sarno, M.~Schmeling, and A.~Schrimpf.
\newblock {C}arry {T}rades and {G}lobal {F}oreign {E}xchange {V}olatility.
\newblock {\em Journal of Finance}, 67(2):681--718, April 2012.

\bibitem{musiela2011martingale}
M.~Musiela and M.~Rutkowski.
\newblock {\em {M}artingale {M}ethods in {F}inancial {M}odelling}, volume~36.
\newblock Springer, 2011.

\bibitem{Nelsen2006}
R.~B. Nelsen.
\newblock {\em {A}n {I}ntroduction to {C}opulas}.
\newblock Springer, 2006.

\end{thebibliography}
